\documentclass[12pt]{article}

\usepackage{amssymb}
\usepackage{amsmath}

\usepackage{epsfig}

\begin{document}%

\title{\bf RS model with a small curvature and Drell-Yan process at the LHC}

\author{A.V. Kisselev\thanks{Presented at the CMS Workshop
on Perspectives on Physics and on CMS at Very High
Luminosity, Alushta, Crimea, Ukraine, 28-31 May, 2012} \\
\small Institute for High Energy Physics, 142281 Protvino, Russia}

\date{}

\maketitle

\thispagestyle{empty}


\begin{abstract}
The $p_{\perp}$-distribution for a dilepton  production at the LHC
with high luminosity is calculated in the Randall-Sundrum scenario
with a small curvature $\kappa$. The widths of massive gravitons are
taken into account. The discovery limits on 5-dimensional gravity
scale $M_5$ are obtained to be 17.8 TeV and 21.6 TeV for $\sqrt{s} =
14$ TeV and integrated luminosities 1000 fb$^{-1}$ and 3000
fb$^{-1}$, respectively. Contrary to the standard RS model, these
limits do not depend on parameter $\kappa$.
\end{abstract}


\section{Introduction}
\label{sec1}

The Randall-Sundrum (RS) model~\cite{Randall:99} is a theory with
one extra dimension (ED) in a slice of the AdS$_5$ space-time. In
the present paper we will study the RS scenario with the
5-dimensional Planck scale $M_5$ in the TeV region and small
curvature~\cite{Giudice:05, Kisselev:06} (se also
Refs.~\cite{Kisselev:ED}):
\begin{equation}\label{scale_relation}
\kappa \ll \bar{M}_5 \;.
\end{equation}
It has the following background warped metric:
\begin{equation}\label{metric}
ds^2 = e^{2 \kappa (\pi r_c - |y|)} \, \eta_{\mu \nu} \, dx^{\mu} \,
dx^{\nu} + dy^2 \;,
\end{equation}
where $y = r_c \, \theta$ ($-\pi \leq \theta \leq \pi$), $r_c$ being
the ``radius'' of the ED, and  $\eta_{\mu \nu}$ is the Minkowski
metric. The points $(x_{\mu}, y)$ and $(x_{\mu}, -y)$ are
identified, so one gets the orbifold $\mathrm{S}^1/\mathrm{Z_2}$.
The parameter $\kappa$ defines a 5-dimensional scalar curvature of
the AdS$_5$ space. In what follows, we will call it
``\emph{curvature}''.

We consider RS scheme with two 3D branes located in 5-th dimension
at the points $y = \pi r_c$ (TeV brane) and  $y = 0$ (Plank brane).
If $\kappa > 0$, the tension on the TeV brane is negative, whereas
the tension on the Planck brane is positive. The SM fields are
constrained to the TeV brane, while the gravity propagates in all
spatial dimensions.

It is necessary to note that metric (\ref{metric}) is chosen in such
a way that 4-dimensional coordinates $x^{\mu}$ are Galilean on the
TeV brane where all the SM fields live, since the warp factor is
equal to unity at $y = \pi r_c$.

By integrating 5-dimensional action in variable $y$, one gets an
effective 4-dimensional action which, in its turn, leads to the
so-called ``hierarchy relation'' between the 5-dimensional reduced
Planck mass $\bar{M}_{\mathrm{Pl}}$ and 5-dimensional reduced
gravity scale $\bar{M}_5$:
\begin{equation}\label{RS_hierarchy_relation}
\bar{M}_{\mathrm{Pl}}^2 = \frac{\bar{M}_5^3}{\kappa} \left( e^{2 \pi
\kappa r_c} - 1 \right) \;.
\end{equation}
Let us notice that $\bar{M}_5$ and 5-dimensional gravity scale $M_5$
are related as follows:
\begin{equation}\label{reduced_scale}
M_5 = (2\pi)^{1/3} \, \bar{M}_5 \simeq 1.84 \, \bar{M}_5 \;.
\end{equation}
In order the hierarchy relation (\ref{RS_hierarchy_relation}) to be
satisfied, it is enough to take $r_c \kappa \simeq 8 \div 9.5$ that
corresponds to $r_c \simeq 0.15 \div 1.8 \mathrm{\ fm}$. Thus, no
large scales (of the order of the Planck mass) are introduced.

From the point of view of a 4-dimensional observer located on the
TeV brane, in addition to the massless graviton, there exists an
infinite number of its Kaluza-Klein (KK) excitations, $G^{(n)}_{\mu
\nu}$, with the masses
\begin{equation}\label{graviton_masses}
m_n = x_n \, \kappa, \qquad n=1,2 \ldots \;,
\end{equation}
where $x_n$ are zeros of the Bessel function $J_1(x)$. Note that
$x_n \simeq \pi (n + 1/4)$ at large $n$.

The interaction of the KK gravitons with the the SM fields
 on the TeV brane is described by the Lagrangian:
\begin{equation}\label{Lagrangian}
\mathcal{L}_{int} = - \frac{1}{\bar{M}_{\mathrm{Pl}}} \, T^{\mu \nu}
\, G^{(0)}_{\mu \nu} - \frac{1}{\Lambda_{\pi}} \, T^{\mu \nu} \,
\sum_{n=1}^{\infty} G^{(n)}_{\mu \nu} \;.
\end{equation}
Here $T^{\mu \nu}$ is the energy-momentum tensor of the matter, and
$n$ is the KK-number. The parameter
\begin{equation}\label{lambda}
\Lambda_{\pi} = \bar{M}_5 \,\left( \frac{\bar{M}_5}{\kappa}
\right)^{\! 1/2}
\end{equation}
is the physical scale on the TeV brane.

In a number of papers, the graviton contribution in the Drell-Yan
process was studied in the standard RS model~\cite{Randall:99}, in
which $\kappa \sim \bar{M}_5 \sim \bar{M}_{\mathrm{Pl}}$ (see, for
instance, \cite{Osland:10}). In such a scheme, an experimental
signature is a real or virtual production of massive KK graviton
resonances. The bounds on the fundamental gravity scale and/or mass
of the lightest resonance were obtained both at the
Tevatron~\cite{Tevatron_limits} and recently at the
LHC~\cite{LHC_limits}. Note that all these bounds significantly
depend on a value of another parameter of the model $\kappa =
0.01\div0.1$.

The RS model with the small curvature has been checked by the DELPHI
Collaboration~\cite{LEP_limit} by studying photon energy spectrum in
the process $e^+e^- \rightarrow \gamma +$ \emph{missing energy}. The
following bound was obtained:
\begin{equation}\label{LEP_limit}
\bar{M}_5 > 0.92 \mathrm{\ TeV}\;.
\end{equation}

There are several reasons to consider RS scenario with the small
curvature:
\begin{itemize}
\item
There exists a serious shortcoming of the standard scenario with the
curvature and fundamental gravity scale being of the order of the
Planck mass, $\kappa \sim \bar{M}_5 \sim
\bar{M}_{\mathrm{Pl}}$~\cite{Boos:02}.
\item
The spectrum of the KK gravitons (\ref{graviton_masses}) is very
similar to that in the model with \emph{one}
ED~\cite{Arkani-Hamed:98}. Note that matrix elements for the
scattering of the SM fields can be formally obtained from
corresponding matrix elements calculated in the model with one
\emph{flat} dimension by using the following
replacement~\cite{Kisselev:06}:
\begin{equation}\label{RS_vs_ADD}
\bar{M}_{4+1} \rightarrow (2 \pi)^{-1/3} \, \bar{M}_5 \;, \qquad R_c
\rightarrow (\pi \kappa)^{-1} \;.
\end{equation}
Here $\bar{M}_{4+1}$ is a 5-dimensional reduced
Planck scale, $R_c$ being the radius of the extra flat dimension. As
a result, all cross sections appear to be rather large (as in the
ADD model with $4+1$ dimensions).
\item
At the same time, astrophysical restrictions are not applied to the
RS-like model with the small curvature, contrary to the ADD model
with one or two EDs~\cite{Kisselev:06}.
\end{itemize}


\section{KK graviton contribution to dilepton production}
\label{sec2}

Let us consider the dilepton production~\cite{Matveev:69, Drell:70}
with high transverse momenta,
\begin{equation}\label{process}
p \, p \rightarrow l^+ l^- + X \;,
\end{equation}
where $l = e \mathrm{\ or\ } \mu$.

The differential cross section of this process is equal to
\begin{eqnarray}\label{cross_sec}
\frac{d \sigma}{d p_{\perp}}(p p \rightarrow  l^+ l^- + X) &=& \!\!
2p_{\perp} \!\!\! \sum\limits_{a,b = q,\bar{q},g} \int\nolimits
d\tau \frac{\sqrt{\tau}}{\sqrt{\tau - x_{\perp}^2}} \int\nolimits \!
\frac{dx_a}{x_a} f_{a/p}(\mu^2, x_a)
\nonumber \\
&\times& \!\!  f_{b/p}(\mu^2, \tau/x_a) \, \frac{d
\hat{\sigma}}{d\hat{t}}(a b \rightarrow  l^+ l^-) \;,
\end{eqnarray}
where $p_{\perp}$ is the transverse momenta of the leading lepton,
and $x_{\perp} = 2p_{\perp}/\sqrt{s}$.

The contribution of the virtual gravitons to the process
(\ref{process}) comes from the quark-antiquark annihilation, $q \,
\bar{q} \rightarrow G^{(n)} \rightarrow l^+l^-$, and gluon-gluon
fusion, $g \, g \rightarrow G^{(n)} \rightarrow l^+l^-$. The
corresponding partonic cross sections are~\cite{Giudice:05}:
\begin{eqnarray}
\frac{d\hat{\sigma}}{d\hat{t}}(q \bar{q} \rightarrow l^+l^-) &=&
\frac{\hat{s}^4 + 10\hat{s}^3 \hat{t} + 42 \hat{s}^2 \hat{t}^2 + 64
\hat{s} \hat{t}^3 + 32\hat{t}^4}{1536\pi \hat{s}^2} \left|
\mathcal{S}(\hat{s}) \right|^2 \;,
\label{quark_cross_sec} \\
\frac{d\hat{\sigma}}{d\hat{t}}(gg \rightarrow l^+l^-)
&=&
-\frac{\hat{t}(\hat{s} + \hat{t}) (\hat{s}^2 + 2 \hat{s} \hat{t} +
2\hat{t}^2)}{256 \pi \hat{s}^2} \left|\mathcal{S}(\hat{s}) \right|^2
 \;, \label{gluon_cross_sec}
\end{eqnarray}
where
\begin{equation}\label{KK_sum}
\mathcal{S}(\hat{s}) = \frac{1}{\Lambda_{\pi}^2} \sum_{n=1}^{\infty}
\frac{1}{\hat{s} - m_n^2 + i \, m_n \Gamma_n} \;.
\end{equation}
Here $\Gamma_n$ denotes a total width of the graviton with the KK
number $n$ and mass $m_n$:
\begin{equation}\label{graviton_widths}
\Gamma_n = \eta \, m_n \left( \frac{m_n}{\Lambda_{\pi}} \right)^2
\;,
\end{equation}
with $\eta \simeq 0.09$~\cite{Kisselev:06}.

In the RS scenario with the small curvature, sum (\ref{KK_sum}) was
calculated analytically in Ref.~\cite{Kisselev:06}:
\begin{equation}\label{grav_propagator}
\mathcal{S}(\hat{s}) = - \frac{1}{4 \bar{M}_5^3 \sqrt{\hat{s}}} \;
\frac{\sin 2A + i \sinh 2\varepsilon }{\cos^2 \! A + \sinh^2 \!
\varepsilon } \;,
\end{equation}
where
\begin{equation}\label{parameters}
A = \frac{\sqrt{\hat{s}}}{\kappa} \;, \qquad \varepsilon =
\frac{\eta}{2} \Big( \frac{\sqrt{\hat{s}}}{\bar{M}_5} \Big)^3 \;.
\end{equation}

\begin{figure}[hbtp]
 \begin{center}
    \resizebox{11cm}{!}{\includegraphics{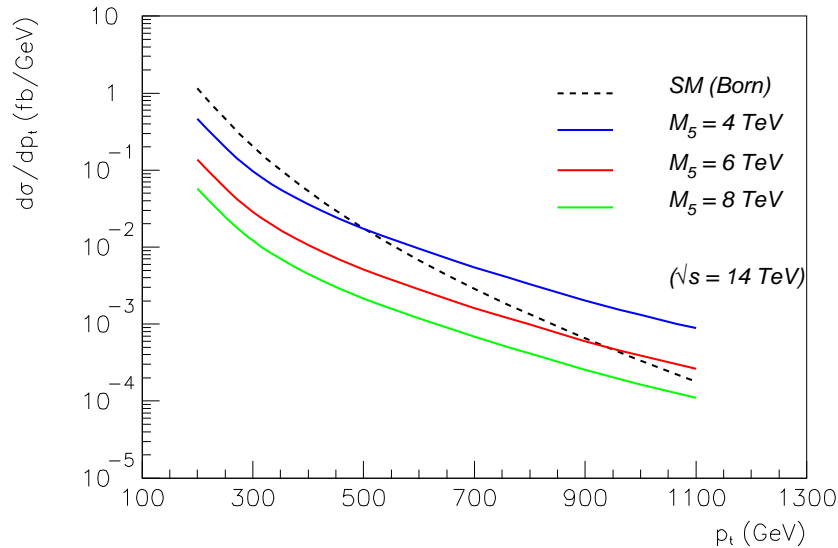}}
\caption{The KK graviton contribution to the Drell-Yan process for
several values of 5-dimensional \emph{reduced} Planck scale (solid
curves) vs. SM contribution (dashed curve) at the LHC.}
    \label{fig:DY_cs_14_TeV}
  \end{center}
\end{figure}

In order to obtain search limits for the LHC, we have calculated
contributions of $s$-channel gravitons to $p_{\perp}$-distributions
of the final leptons for different values of $\bar{M}_5$ (see
Fig.~\ref{fig:DY_cs_14_TeV}). We used the MSTW 2008 NNLO parton
distributions~\cite{MSTW}, and convolute them with the partonic
cross sections (\ref{quark_cross_sec}), (\ref{gluon_cross_sec}) in
Eq.~(\ref{cross_sec}). The PDF scale was taken to be equal to the
invariant mass of the final leptons, $\mu = \sqrt{\hat{s}}$. The
following limit on pseudorapidity of the final leptons was imposed:
\begin{equation}\label{rapidity_cut}
|\eta| \leq 2.5 \,.
\end{equation}
It defines a region of integration in variables $x_a$, $\tau$ in
(\ref{cross_sec}). The reconstruction efficiency of 90$\%$ was
assumed for both electrons and muons~\cite{Efficiency}.

Let us stress that in our scheme the ``new physics'' contribution
\emph{do not depend} on $\kappa$, provided $\kappa \ll M_5$, in
contrast to the RS model with the large curvature
$\kappa$~\cite{Randall:99}.

\begin{figure}[hbtp]
 \begin{center}
    \resizebox{11cm}{!}{\includegraphics{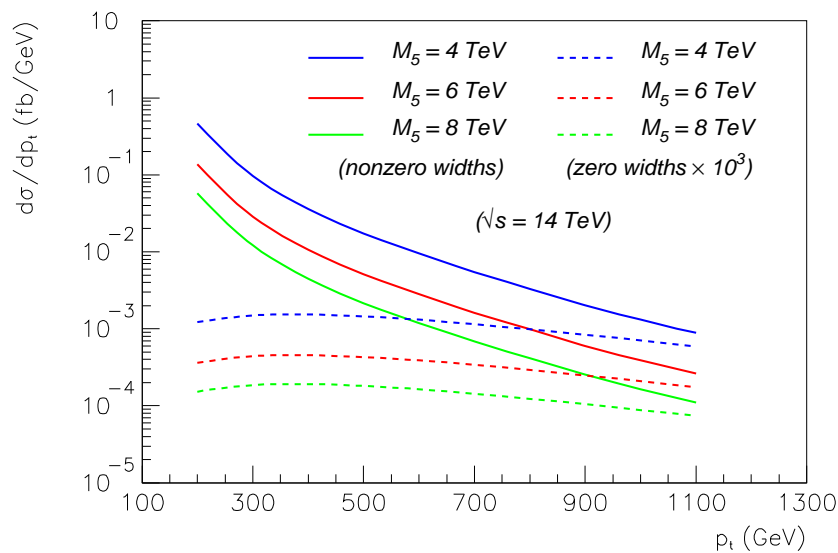}}
\caption{Contributions to the dilepton production from KK gravitons
with nonzero widths (solid curves) vs. contributions from zero width
graviton (dashed curves) for different values of the \emph{reduced}
gravity scale at the LHC.}
    \label{fig:DY_cs_14_zero_TeV}
  \end{center}
\end{figure}

Note that an ignorance of the graviton widths would be a \emph{rough
approximation}. In such a case, we get from (\ref{KK_sum}):
\begin{equation}\label{KK_sum_zero_widths}
\mathrm{Im} \,\mathcal{S}(\hat{s}) = - \frac{1}{2 \bar{M}_5^3
\sqrt{\hat{s}}} \;, \qquad  \mathrm{Re} \, \mathcal{S}(\hat{s}) = 0
\;.
\end{equation}
Fig.~\ref{fig:DY_cs_14_zero_TeV} shows the graviton contribution to
the $p_{\perp}$-distribution calculated with the use of
Eqs.~(\ref{KK_sum_zero_widths}). As one can see, neglecting graviton
widths results in very large suppression of the cross sections.

\begin{figure}[hbtp]
 \begin{center}
    \resizebox{11cm}{!}{\includegraphics{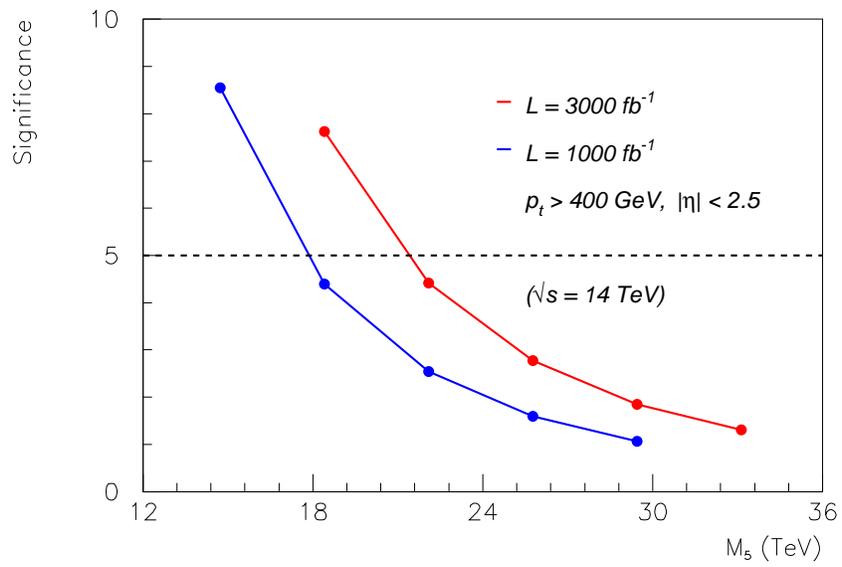}}
\caption{The statistical significance for the process $p \, p
\rightarrow l^+ l^- + X $ at the LHC with high integrated luminosity
as a function of the 5-dimensional Planck scale $M_5$.}
    \label{fig:DY_S_14_TeV_hl}
  \end{center}
\end{figure}

To evaluate the event rates, we used a K-factor 1.5 for the SM
background, while a conservative value of K=1 was taken for the
signal. Only \emph{one} type of leptons (electron or muon) was taken
into account. Let $N_S$ ($N_B$) be a number of signal (background)
events with $p_{\perp} > 400$ GeV. Then we define the statistical
significance $\mathcal{S} = N_S/\sqrt{N_B}$, and require a $5
\sigma$ effect. A statistical significance is presented in
Fig.~\ref{fig:DY_S_14_TeV_hl} as a function of the 5-dimensional
Planck scale $M_5$ for two values of an integrated luminosity.
Remember that $M_5$ and $\bar{M}_5$ are related by
Eq.~(\ref{reduced_scale}).

\section{Conclusions}
\label{sec3}

We have considered the dilepton production in the scenario with one
warped extra dimension and small curvature
$\kappa$~\cite{Giudice:05, Kisselev:06}. In such a scheme, the
reduced 5-dimensional Planck scale $\bar{M}_5$ can vary from one to
tens TeV, while $\kappa \ll \bar{M}_5$. The mass spectrum is similar
to that in the ADD model~\cite{Arkani-Hamed:98} with one flat extra
dimension.

The $p_{\perp}$-distributions for the Drell-Yan process, $p p
\rightarrow l^+ l^- + X$ ($l=e$, or $\mu$), are calculated for the
collision energy $\sqrt{s} = 14$ TeV. The following LHC discovery
limits are obtained:
\begin{equation}\label{search_limit}
M_5 = \left\{
\begin{array}{rl}
  17.8 \mathrm{\ TeV} \;, & \mathcal{L} = 1000 \mathrm{\
fb}^{-1} \\
  21.6 \mathrm{\ TeV} \;,  & \mathcal{L} = 3000 \mathrm{\
fb}^{-1} \\
\end{array}
\right.
\end{equation}
It is important to note that these bounds on $M_5$ do not depend on
$\kappa$, contrary to the standard RS model with the large
curvature~\cite{Randall:99}.



\end{document}